\documentclass[apj]{emulateapj}

\shorttitle{The Highly-Dynamic Cluster Core of A1664}
\shortauthors{Calzadilla et al.}

\usepackage{amssymb,amsmath}
\usepackage{graphicx}
\usepackage{appendix}
\usepackage{natbib}
\usepackage[backref,breaklinks,colorlinks,citecolor=blue,linkcolor=blue]{hyperref}

\usepackage{color}

\usepackage{stackengine}

\def\MIT{1}
\def\IoA{2}
\def\UManitoba{3}
\def\RIT{4}
\def\LERMA{5}
\def\MSU{6}
\def\Durham{7}
\def\UWaterloo{8}
\def\PI{9}
\def\CfA{10}
\def\ICRAR{11}
\def\SURFsara{12}
\def\Leiden{13}
\def\ASTRON{14}
\def\Yale{15}

\begin{document}

\title{Revealing a Highly-Dynamic Cluster Core in Abell 1664 with \emph{Chandra}}

\author{
Michael S. Calzadilla\altaffilmark{\MIT,\IoA},
Helen R. Russell\altaffilmark{\IoA},
Michael A. McDonald\altaffilmark{\MIT},
Andrew C. Fabian\altaffilmark{\IoA},\\
Stefi A. Baum\altaffilmark{\UManitoba,\RIT},
Fran\c{c}oise Combes\altaffilmark{\LERMA},
Megan Donahue\altaffilmark{\MSU},
Alastair C. Edge\altaffilmark{\Durham},
Brian R. McNamara\altaffilmark{\UWaterloo,\PI},\\
Paul E. J. Nulsen\altaffilmark{\CfA,\ICRAR},
Christopher P. O'Dea\altaffilmark{\UManitoba,\RIT},
J. B. Raymond Oonk\altaffilmark{\SURFsara,\Leiden,\ASTRON},\\
Grant R. Tremblay\altaffilmark{\CfA,\Yale},
Adrian N. Vantyghem\altaffilmark{\UWaterloo}
}

\affiliation{
$^{\MIT}${Kavli Institute for Astrophysics and Space Research, Massachusetts Institute of Technology, Cambridge, MA 02139, USA}\\
$^{\IoA}${Institute of Astronomy, Madingley Road, Cambridge CB3 0HA, UK}\\
$^{\UManitoba}${University of Manitoba, Dept. of Physics and Astronomy, Winnipeg, MB R3T 2N2, Canada}\\
$^{\RIT}${School of Physics \& Astronomy, Rochester Institute of Technology, 84 Lomb Memorial Dr., Rochester, NY 14623, USA}\\
$^{\LERMA}${LERMA, Observatoire de Paris, PSL Research Univ., Coll\`{e}ge de France, CNRS, Sorbonne Univ., UPMC, Paris, France}\\
$^{\MSU}${Physics and Astronomy Dept., Michigan State University, East Lansing, MI, 48824 USA}\\
$^{\Durham}${Centre for Extragalactic Astronomy, Department of Physics, Durham University, Durham, DH1 3LE, UK}\\
$^{\UWaterloo}${Department of Physics \& Astronomy, University of Waterloo, 200 University Avenue, West Waterloo, Ontario N2L 3G1, Canada}\\
$^{\PI}${Perimeter Institute for Theoretical Physics, 31 Caroline Street North, Waterloo, ON N2L 2Y5, Canada}\\
$^{\CfA}${Harvard-Smithsonian Center for Astrophysics, 60 Garden Street, Cambridge, MA 02138, USA}\\
$^{\ICRAR}${ICRAR, University of Western Australia, 35 Stirling Highway, Crawley, WA 6009, Australia}\\
$^{\SURFsara}${SURFsara, P.O. Box 94613, 1090 GP Amsterdam, the Netherlands}\\
$^{\Leiden}${Leiden Observatory, Leiden University, P.O. Box 9513, NL-2300 RA Leiden, The Netherlands}\\
$^{\ASTRON}${Netherlands Institute for Radio Astronomy (ASTRON), Postbus 2, 7990 AA Dwingeloo, The Netherlands}\\
$^{\Yale}${Yale Center for Astronomy and Astrophysics, Yale University, 52 Hillhouse Ave., New Haven, CT 06511, USA}
}

\email{msc92@mit.edu}

\begin{abstract}

We present new, deep (245 ks) {\it Chandra} observations of the galaxy cluster Abell 1664 ($z = 0.1283$). These images reveal rich structure, including elongation and accompanying compressions of the X-ray isophotes in the NE-SW direction, suggesting that the hot gas is sloshing in the gravitational potential. This sloshing has resulted in cold fronts, at distances of 55, 115 and 320 kpc from the cluster center. Our results indicate that the core of A1664 is highly disturbed, as  
the global metallicity and cooling time flatten at small radii, implying mixing on large scales. The central AGN appears to have recently undergone a mechanical outburst, as evidenced by our detection of cavities. These cavities are the X-ray manifestations of radio bubbles inflated by the AGN, and may explain the motion of cold molecular CO clouds previously observed with {\it ALMA}. The estimated mechanical power of the AGN, using the minimum energy required to inflate the cavities as a proxy, is $P_{cav} = (1.1 \pm 1.0) \times 10^{44} $ erg s$^{-1}$, which may be enough to drive the molecular gas flows, and offset the cooling luminosity of the ICM, at $L_{\rm cool} = (1.90 \pm0.01)\times 10^{44}$ erg s$^{-1}$. This mechanical power is orders of magnitude higher than the measured upper limit on the X-ray luminosity of the central AGN, suggesting that its black hole may be extremely massive and/or radiatively inefficient. We map temperature variations on the same spatial scale as the molecular gas, and find that the most rapidly cooling gas is mostly coincident with the molecular gas reservoir centered on the BCG's systemic velocity observed with {\it ALMA} and may be fueling cold accretion onto the central black hole. 

\end{abstract}





\section{Introduction}
\label{sec:intro}

Galaxy clusters can contain hundreds to thousands of galaxies, and usually have on the order of $10^{13}$--$10^{14}$ M$_{\odot}$ of hot gas called the intracluster medium (ICM), which comprises $\sim 10 \%$ of the total cluster mass and is observable in X-rays, and permeates the space between the member galaxies.
In such gas-rich systems, we should observe runaway cooling flows at low temperatures which, if unhindered, would fuel cooling flows of 100--1000 M$_{\odot}$ yr$^{-1}$, which in turn should lead to cold gas reservoirs of 5 -- 50 $\times ~ 10^{11} $ M$_{\odot}$ \citep[e.g.][]{1994ARA&A..32..277F}. Instead, UV, optical, and infrared observations of the central brightest cluster galaxy (BCG) reveal highly-suppressed star formation rates of 1 -- 100 M$_{\odot}$ yr$^{-1}$ \citep[e.g.][]{1987MNRAS.224...75J,1987ApJ...323L.113R,1989AJ.....98.2018M,1993MNRAS.265..431C,1995MNRAS.276..947A,1999MNRAS.306..857C,2001A&A...365L..93M,2004ApJ...601..173M,2006ApJ...652..216R,2008ApJ...681.1035O,2015ApJ...805..177D,2015MNRAS.450.2564M,2018ApJ...858...45M}. Adding to this mystery, high-resolution X-ray spectra of cooling flows revealed that many of the characteristic recombination lines in the cooling gas were much weaker than expected, consistent with cooling being suppressed by 1--2 orders of magnitude (\citealt{2003ApJ...590..207P,2006PhR...427....1P,2010A&ARv..18..127B} for a review). 

It has become clear from these and other works that the ICM is not cooling unimpeded. Feedback from active galactic nuclei (AGN) is almost certainly responsible for the discrepancy in predicted versus observed cooling levels \citep{2007ARA&A..45..117M,2012NJPh...14e5023M,2012ARA&A..50..455F}. At low accretion rates, AGN operate in a radiatively inefficient (radio or kinetic) mode, where most of their energy output is mechanical in the form of powerful radio jets \citep{2005MNRAS.363L..91C}. As these jets expand into large radio-emitting lobes, the ICM is displaced by these bubbles, creating visible cavities in X-ray images, and high spatial resolution instruments like {\it Chandra } have shown these to be energetically capable of preventing large-scale cooling \citep[e.g.][]{2007ARA&A..45..117M}. By measuring the extent of these bubbles, we can estimate the heat input by the jets from the mechanical $pV$ work needed to inflate the bubbles against their surrounding gas pressure \citep{2002MNRAS.332..729C}. The mean jet power is comparable to the rate of cooling, and therefore able to quench cooling in a moderated feedback loop \citep[e.g.][]{2004ApJ...607..800B,2006MNRAS.373..959D,2006ApJ...652..216R}. It is still unclear how exactly this energy couples to the surrounding hot atmosphere; some possible coupling mechanisms include weak shocks and sound waves \citep[e.g.][]{2003MNRAS.344L..43F,2007MNRAS.381.1381S,2017MNRAS.464L...1F}, turbulence induced by g-modes \citep[e.g.][]{2011MNRAS.414.1493R,2018ApJ...857...84B}, turbulent dissipation and mixing \citep[e.g.][]{2014Natur.515...85Z,2002MNRAS.332..729C,2003ApJ...596L.139K}, and cosmic rays \citep[e.g.][]{2006ApJ...642..140C}.

AGN feedback is included in many galaxy formation simulations \citep[e.g.][]{2006MNRAS.365...11C,2006MNRAS.370..645B,2014MNRAS.444.1518V}, though we still do not understand how outbursts work on local scales, close to the AGN, and how these small-scale energetics couple and transport energy to large scales and suppress cooling. In clusters, molecular gas and star formation are preferentially observed as the central cooling time falls below a remarkably sharp threshold of $5 \times 10^8$ yr, or the entropy below 30 keV cm$^2$ \citep{2008ApJ...687..899R,2008ApJ...683L.107C}. Multi-wavelength studies throughout the years have been able to link AGN feedback to the presence of this molecular gas in cluster cores \citep[e.g.][]{1994ApJ...422..467O,2001MNRAS.328..762E,2003MNRAS.344L..48F,2005MNRAS.363..216C,2005MNRAS.361...17C,2005MNRAS.360..748J,2006MNRAS.367..433H,2008ApJ...672..252L,2010ApJ...721.1262M,2010MNRAS.405..898O,2013MNRAS.435.1108C}. More recently, early results from the Atacama Large Millimeter Array (\emph{ALMA}), with its unprecedented spatial resolution at sub-mm wavelengths, have shown that molecular gas flows have relatively low velocities, and are morphologically coupled to bubbles \citep[e.g.][]{2016MNRAS.458.3134R,2017ApJ...836..130R}. It is possible that a fraction of this cool gas could serve as fuel for the central SMBH as it condenses out of the hot cluster halo, thereby coupling the AGN's accretion to the large-scale cooling rate \citep[e.g.][]{2010MNRAS.408..961P,2012ApJ...746...94G}.

Feedback from the AGN, whether mechanical or radiative, is not the only way to offset cooling. There are additional processes that serve to mix or heat the ICM. For instance, early {\it Chandra} observations revealed the presence of cold fronts in relaxed galaxy clusters, which are sharp contact discontinuities between gases of different temperatures and densities. \citet{2006ApJ...650..102A} concluded that these cold fronts are caused by infalling subhalos being stripped of their gas early on. As a result of the changing shape of the gravitational potential, the cluster core then oscillates and causes changes in ram pressure, giving the infalling gas angular momentum and resulting in cold fronts in a characteristic spiral pattern about the core \citep[see][for a review]{2007PhR...443....1M}. These major and minor galaxy mergers can mix low- and high-entropy gas \citep[e.g.][]{2010ApJ...717..908Z}, but the resulting sloshing may not operate on short enough timescales to prevent runaway cooling flows.

\begin{table}
\centering
\caption{Summary of {\it Chandra} observations for A1664 \label{tab:observations}}
\begin{tabular}{lccccc}
\hline
ObsID & Date & Instrument  & Exposure  & Cleaned\\
\hline
1648        & 2001-06-08    & ACIS-S      & 9.78   ks     & 9.27 ks \\
7901        & 2006-12-04    & ACIS-S      & 36.56 ks     & 35.98 ks\\
17172       & 2014-12-07    & ACIS-S      & 67.14 ks     & 62.54 ks\\
17173       & 2015-03-14    & ACIS-S      & 19.07 ks     & 18.82 ks\\
17557       & 2014-12-12    & ACIS-S      & 66.74 ks     & 64.58 ks\\
17568       & 2015-03-10    & ACIS-S      & 46.19 ks     & 43.64 ks\\
\hline
\multicolumn{3}{l}{Total:}                & 245.48 ks    & 234.83 ks        \\  
\hline
\hline
\end{tabular}
\end{table}

Here we aim to investigate the effects of AGN feedback and gas sloshing in a nearby galaxy cluster, Abell 1664 (hereafter A1664), a cool core cluster at a redshift of $z=0.1283$ \citep{1992MNRAS.259...67A,1995MNRAS.275..741A}. 
In \autoref{sec:obs} we give details about our observations and how we reduced the data. In \autoref{sec:results} we present our results, then provide an interpretation of them in \autoref{sec:discuss}. Finally, we give a summary of our work in \autoref{sec:end}. For this study, we assume a $\Lambda$CDM cosmology with $H_0$ = 70 km s$^{-1}$ Mpc$^{-1}$, $\Omega_m$ = 0.3 and $\Omega_\Lambda$ = 0.7, which gives an angular scale of 2.291 kpc arcsec$^{-1}$ at the cluster redshift. All errors are 1$\sigma$ unless noted otherwise.


\section{Observations \& Data Reduction}
\label{sec:obs}

\subsection{Chandra X-ray Observations}

\begin{figure*}
\centering
\stackinset{r}{0.6in}{t}{0.15in}{\includegraphics[height=3.0cm]{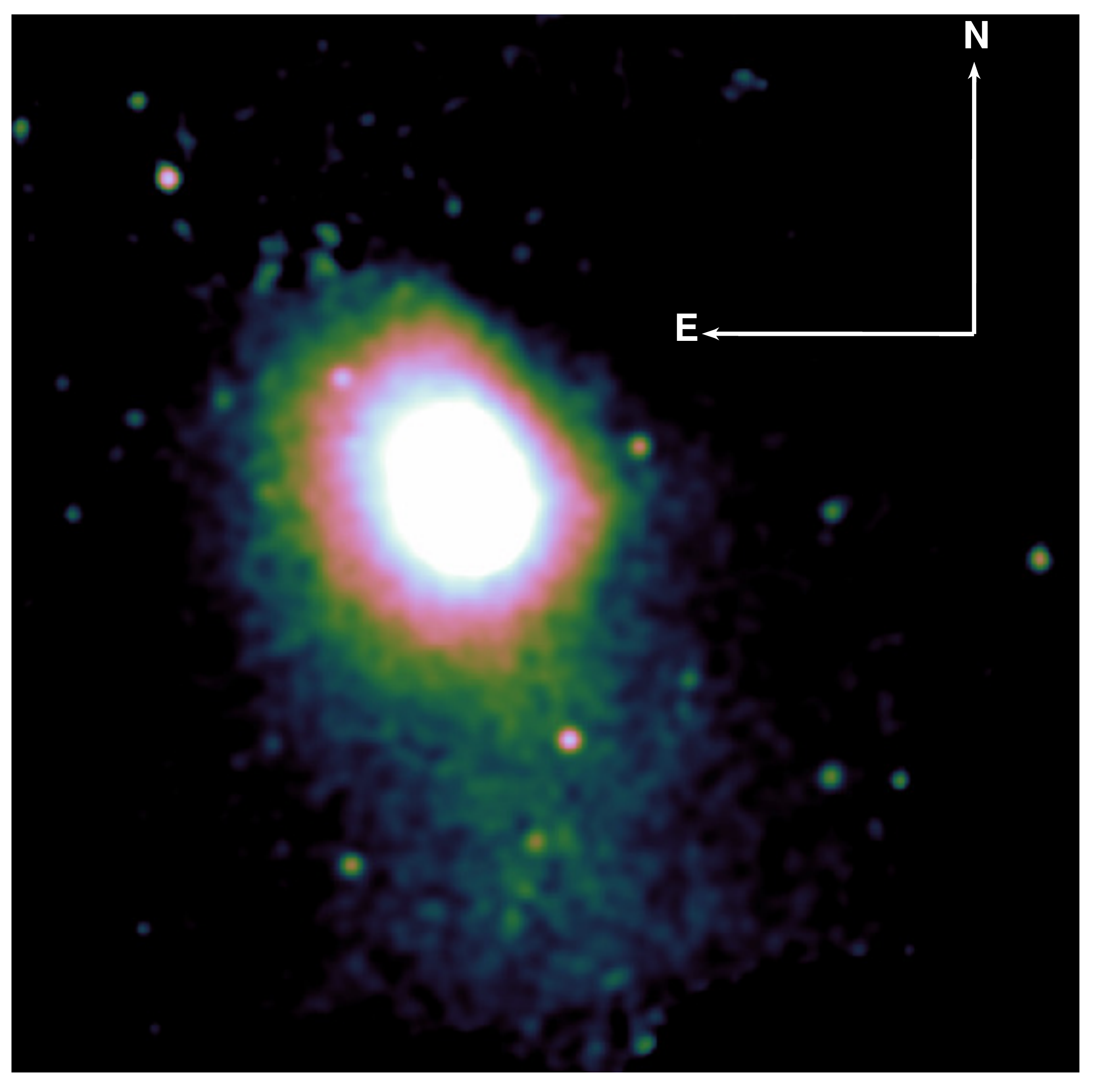}}{\includegraphics[width=0.8\textwidth]{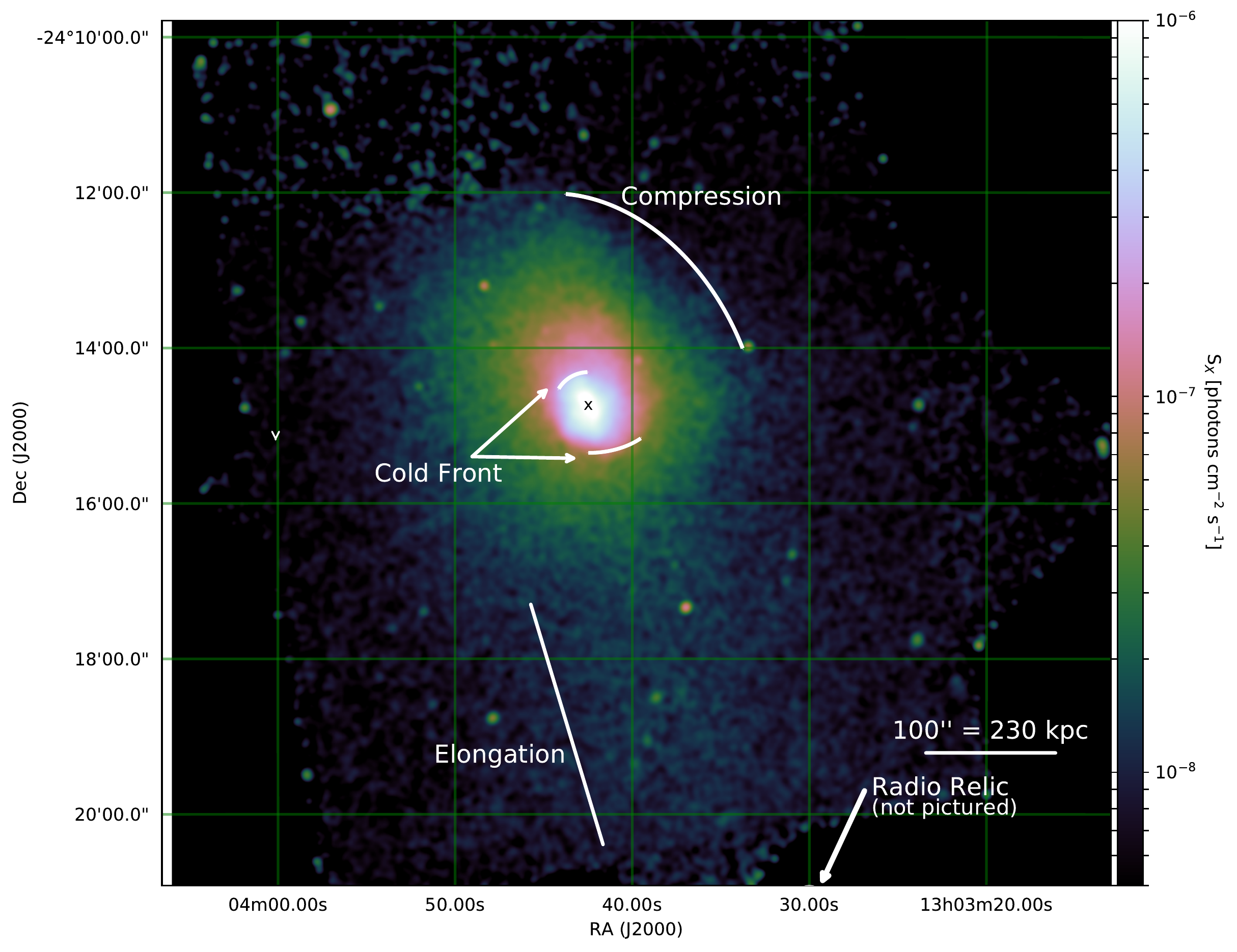}}
\caption{Exposure-corrected {\it Chandra} flux image of A1664, in the 0.5--7.0 keV band, in units of photons cm$^{-2}$ s$^{-1}$. Binning is such that one pixel corresponds to 0.984\arcsec on a side. Logarithmic intensity scaling is used, and the image is smoothed with a 3 pixel wide Gaussian kernel to show the fainter elongation of emission to the South. Compression of the X-ray isophotes is visible 320 kpc ($\sim 140$\arcsec) NW of the cluster center, and again at 115 kpc ($\sim$50\arcsec) to the South and at 55 kpc ($\sim$24\arcsec) to the NE of the center, suggesting N-S gas sloshing on three scales. At the core of the cluster, emission is concentrated in an X-ray bar structure with a NE-SW elongation. The `x' symbol marks the ICM coordinate center at (RA, Dec) = ($13^h03^m42\fs465, -24^{\circ}14'44\farcs671$). The inset to the upper right of the image is a CIAO adaptively smoothed image, created with 5 Gaussian smoothing kernels with a minimum and maximum kernel radius of 1 pixel and 5 pixels respectively, with log spacing between the radii, using the CIAO tool {\it dmimgadapt}. It shows the elongation of emission to the South more clearly. \label{fig:fluximage}}
\end{figure*}

This study introduces four new observations of A1664 (ObsIDs 17172, 17173, 17557, 17568) using the S3 chip of the Advanced CCD Imaging Spectrometer (ACIS) on board the {\it Chandra X-ray Observatory}. Combined with previous observations (ObsIDs 1648, 7901) this analysis makes use of a total exposure time of 245 ks (\autoref{tab:observations}), a 199 ks increase over previous studies \citep[e.g.,][]{2009ApJ...697..867K}. The Chandra Interactive Analysis of Observations (CIAO) software package version 4.8.1 and version 4.7.0 of the calibration database (CALDB) provided by the {\it Chandra} X-ray Center (CXC) were used to reduce the data. The latest gain and charge transfer inefficiency (CTI) corrections were also applied to reprocess the level 1 event files. All of the observations were taken in the {\scriptsize VFAINT} data mode, so improved background screening was applied in making the new level 2 event files using {\it acis\_process\_events}. Background flares were then eliminated using the {\scriptsize LC\_CLEAN} script provided by M. Markevitch, and ended up with a total cleaned exposure of 234.83 ks. Finally, the Blank-sky background files were used for background subtraction, and their exposures were normalized to the count rate of their respective foreground observations in the 9--12 keV band, where \emph{Chandra's} effective area is too low to typically detect point and extended sources.

The observations were reprojected to a common tangent point and merged together. Exposure maps were also calculated using a monoenergetic distribution of source photons of 2.3 keV, as recommended by the CXC for the broad (0.5 -- 7.0 keV) energy band\footnote{\url{http://cxc.harvard.edu/ciao/why/monochromatic\_energy.html}}. To create the blank-sky background exposure maps, a random arrival time within the exposure time of the background observation was first assigned to each photon in the events list, as these files do not have a time column. The same ratio of observation exposure time to background exposure time was imposed on each observation. Then, the ratio of the background to the observation exposure was calculated for each ObsID, which was multiplied by the observation exposure maps to make background exposure maps and then merged together. 

In this study the {\scriptsize MEKAL} XSPEC thermal spectral model \citep{1993A&AS...97..443K} is used to model the emission of an optically-thin plasma, and {\scriptsize WABS} \citep{1992ApJ...400..699B,1983ApJ...270..119M} to model photoelectric absorption. Abundances were measured assuming the ratios from \citet{1989GeCoA..53..197A} for consistency with previous literature.

\subsection{HST Optical Imaging}
In addition to the X-ray data, we present archival Hubble Space Telescope (HST) data obtained by \citet[][project ID 11230]{2010ApJ...719.1619O}, and retrieved from the Mikulski Archive for Space Telescopes (MAST). The imaging data we present here were observed with the broadband F606W filter on the Wide Field and Planetary Camera 2 (WFPC2). The data were reduced using the standard recalibration pipelines, with individual exposures combined using the {\scriptsize ASTRODRIZZLE}\footnote{\url{http://www.stsci.edu/hst/HST\_overview/drizzlepac}} routine, after removing cosmic ray signatures.

\subsection{JVLA Radio Data}
A1664 was also observed with the Karl G. Jansky Very Large Array (JVLA). The JVLA A-array data were obtained in March 2014 (PI Edge, project ID 14A-280) using the L-band receiver. The data were reduced and calibrated with CASA. The image presented uses only the three highest frequency sub-bands (1.80--1.92 GHz) to ensure the best spatial resolution and least RFI. The recovered beam is elliptical due to the low declination of the source but is sufficient to conclude that the source is unresolved on 2--3\arcsec scales.


\section{Data Analysis \& Results}
\label{sec:results}

An exposure-corrected image combining the six \emph{Chandra} observations is shown in \autoref{fig:fluximage}. The image shows cluster emission elongated Southward toward the edge of the CCD chip, with an accompanying surface brightness edge 320 kpc ($\sim 140$\arcsec) NW of the X-ray centroid (marked with an `x' in \autoref{fig:fluximage}). The inset in the upper right of \autoref{fig:fluximage} is an adaptively smoothed image, created using the CIAO tool {\it dmimgadapt}\footnote{\url{http://cxc.harvard.edu/ciao/ahelp/dmimgadapt.html}}, which highlights the elongation of emission to the South more clearly.

The elongation of emission toward the South also leads down toward a radio relic (beyond the S3 chip), at a distance of 1.1 Mpc ($\sim$8\arcmin) from the cluster center, which is suggestive of merger activity \citep[e.g.][]{2001A&A...376..803G,1999NewA....4..141G}. Radio relics are diffuse radio sources located in the peripheries of clusters with no apparent galaxy counterparts, and have been proposed to either be the sites of merger-induced shock fronts in the ICM, or remnants of radio galaxies. They are also proof that large-scale magnetic fields and relativistic electrons are present in the ICM. The relic in A1664 was resolved by \citet{2012ApJ...744...46K} and studied in detail using the Giant Metrewave Radio Telescope (GMRT).

Since X-ray emission in the ICM is dominated by collisional processes, emission per unit volume is proportional to density squared. Thus, the surface brightness edge at $R = $ 320 kpc ($\sim 140$\arcsec) corresponds to a density discontinuity which may be a cold front produced by sloshing. There are additional surface brightness edges/compressions in the X-ray isophotes closer to the center of the cluster, 115 kpc ($\sim 50$\arcsec) South, and again 55 kpc ($\sim 24$\arcsec) North of the cluster center. We investigate these surface brightness edges further as possible cold fronts in \autoref{subsec:sloshing}. At the core of the cluster, the X-ray emission is concentrated in a ``bar'' structure with elongation in the NE-SW direction. These instances of elongation and compression in the isophotes suggest N-S gas sloshing.

The center, ellipticity and position angle of A1664 on the sky were determined by fitting a two-dimensional beta model\footnote{\url{http://cxc.harvard.edu/sherpa/ahelp/beta2d.html}} to the counts image over a region 600 kpc $\approx$ 260\arcsec ~ in radius which covers most of the field of view of the co-added images. Best fit parameters were first found by using CIAO's \emph{Sherpa} package Monte Carlo optimization method with the Cash statistic \citep{1979ApJ...228..939C} to determine a global minimum in parameter space, followed by a Levenberg-Marquardt simplex minimization to more accurately locate a local minimum. 
The X-ray centroid is found at (RA, Dec) = ($13^h03^m42\fs465, -24^{\circ}14'44\farcs671$), about 5.6 kpc ($2.45$\arcsec) SW from the cluster emission peak, and 6.5 kpc ($2.84$\arcsec) from the BCG, found at (RA, Dec) = ($13^h03^m42\fs540, -24^{\circ}14'42\farcs020$). The X-ray centroid found here is used in all subsequent thermodynamic profiles.

\begin{figure*}
\centering
\includegraphics[height=4.5in,width=\textwidth]{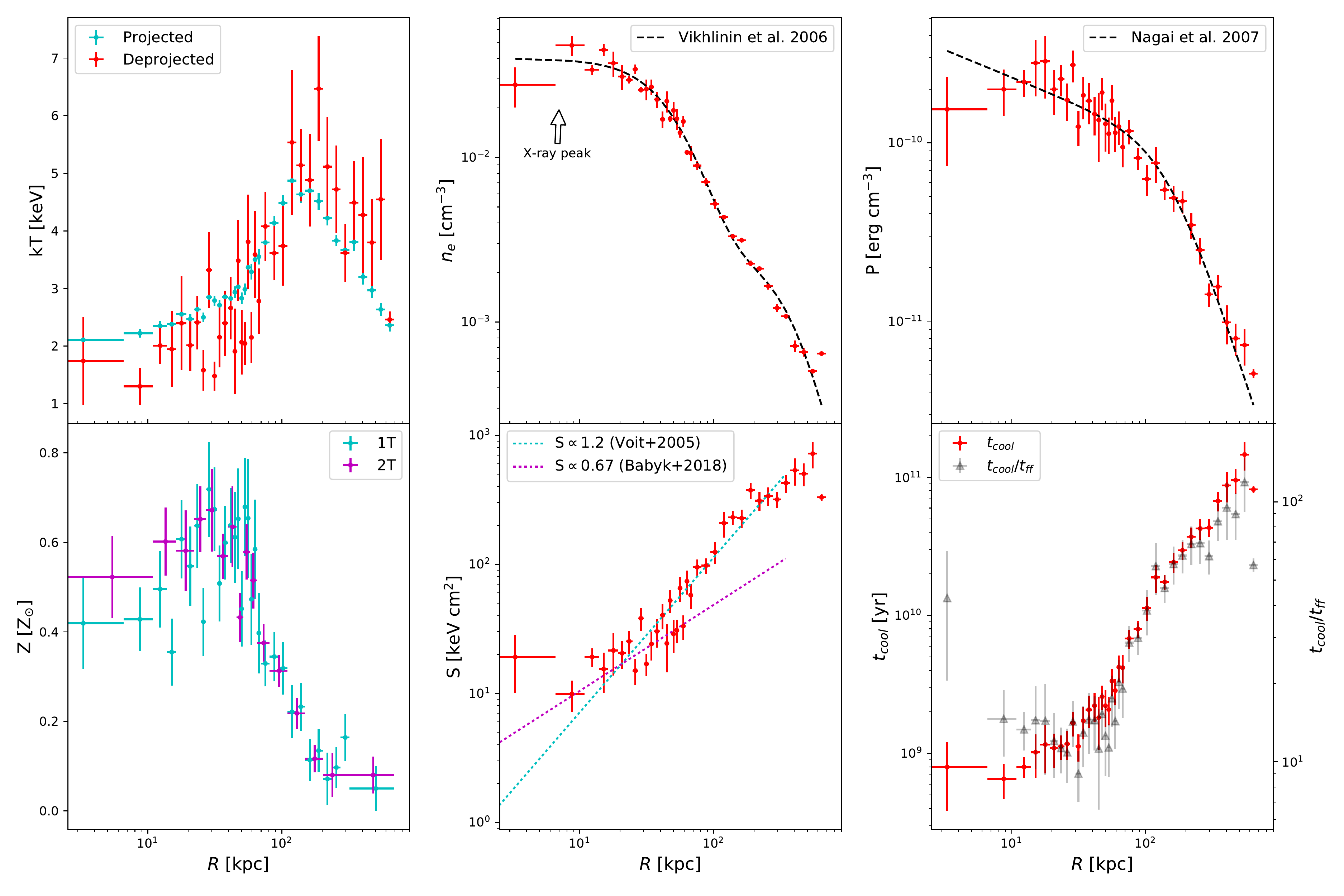}
\caption{\emph{Top left:} Projected (cyan points) vs. deprojected (red points) temperature profile. The factor of $\sim$2 difference in temperature between the maximum (at $R\sim$150 kpc) and minimum (near the core) is characteristic of cool core clusters. \emph{Top center:} deprojected density profile, fit with the universal density profile from \citet{2006ApJ...640..691V}. An arrow indicates the position of the X-ray peak, $\sim 6$ kpc from the X-ray centroid, which makes the central bin seem low. \emph{Top right:} pressure profile, fit with a universal pressure model from \citet{2007ApJ...668....1N}. \emph{Bottom left:} projected metallicity profiles, fit with single- and two-temperature models, calculated with the solar abundance ratios from \citet{1989GeCoA..53..197A}. The flatness in the inner bins implies that these metals are being mixed. \emph{Bottom center:} entropy profile. A power law was fit to find the normalization of the outer radii with a slope of $S \propto r^{1.2}$ following \citet{2005MNRAS.364..909V}. The central bins are not fit well by this power law and are consistent with a second flat or very shallow power law slope following the \citet{2018ApJ...862...39B} $S \propto r^{2/3}$ expectation. \emph{Bottom right:} cooling time profile, with the ratio of the cooling time (based on thermal energy rather than enthalpy) to free fall time profile overlaid. \label{fig:rprof}}
\end{figure*}

\bigskip
\bigskip

\subsection{Global Cluster Properties}

\subsubsection{Total Spectrum}
\label{subsec:spectra}

For the global spectral analysis, the total spectrum was extracted from a high signal-to-noise region (340 kpc = 150\arcsec~ in radius) centered on the cluster core, with point sources excluded, using the CIAO tool {\it dmextract} on the foreground and background events files. Response matrices and ancillary response files for the extracted regions were created using CIAO scripts {\it mkacisrmf} and {\it mkwarf}, respectively, and weighted according to the number of counts in each spectral region. Spectra were extracted separately from each observation then fit jointly over the 0.5 -- 7.0 keV range using XSPEC, version 12.9.0 \citep{1996ASPC..101...17A}.

First and foremost, each of our spectral models were tested once with the Galactic column density ($N_H$) allowed to vary, and again with the $N_H$ fixed to the galactic value of 8.86$\times 10^{20}$ cm$^{-2}$ \citep{2005A&A...440..775K}. Allowing $N_H$ to vary yields better fits, with a chi-squared value of $\chi ^2 = 2239.71$ and 2046 degrees of freedom (as opposed to $\chi ^2 = 2357.11$ with 2047 degrees of freedom), giving an F-test value of 107.246 with a probability of $1.57\times 10^{-24}$. Throughout the rest of this paper, we adopt a new $N_H$ value of 10.81$\times 10^{20}$ cm$^{-2}$, which corresponds to the average value measured in the inner 150\arcsec.

We allowed the {\scriptsize MEKAL} temperature, metallicity, and normalization within the inner $\sim$ 340 kpc (150\arcsec) to vary, with the column densities fixed to $N_H = 10.81 \times 10^{20}$ cm$^{-2}$. The fits reveal that this cluster has an average temperature of 3.58 $\pm$ 0.02 keV, and a metallicity of 0.36 $\pm$ 0.01 Z$_{\odot}$, within a radius of $r<340$ kpc.

\subsubsection{Deprojection of ICM Profiles}

To learn more about how the ICM varies on smaller scales, radial profiles of various quantities were constructed using regions concentric about the X-ray centroid and containing roughly 10,000 counts in the 0.5--7.0 keV band. However, extracting a spectrum from any inner region of the cluster on the plane of the sky will include spectral contributions from regions both closer to or further from us along the line of sight. Therefore, it is necessary to subtract the projected contributions along the line of sight to extract more accurate spectra in the core. It is then necessary to make an assumption about the line-of-sight extent, and a standard way to do this is to assume the emission can be deprojected in a series of concentric spherical shells \citep{1981ApJ...248...47F,1983ApJ...272..439K}. We use the direct spectral deprojection ({\scriptsize DSDEPROJ}) routine from \citet{2007MNRAS.381.1381S} (see also \citealt{2008MNRAS.390.1207R}).

To demonstrate the effects of projection on cluster emission, single-temperature, azimuthally symmetric radial temperature profiles are shown in \autoref{fig:rprof}. The deprojected temperature profile stays relatively close to the projected one, with a minimum value of $\sim$1.5 keV, interior to $\sim$10 kpc. The projected metallicity profile is also shown here. The profile flattens in the center, implying that these metals are being mixed on up to $\sim 70$ kpc (30\arcsec) scales. We note that we expect the ICM to be multiphase at the cluster center, where single-{\scriptsize MEKAL} (1T) fits are sensitive to the Fe-L complex at low $kT$, resulting in a broad spectral shape that gets treated as a lower metallicity feature \citep{2000MNRAS.311..176B}. To investigate this potential bias, we include the metallicity profile after performing a two-{\scriptsize MEKAL} (2T) fit, and find that the profile still flattens at small radii when we account for multiphase structure. The rest of the profiles are only shown in deprojection, in \autoref{fig:rprof}.

The electron density profile was derived from the {\scriptsize MEKAL} normalization in XSPEC by
\begin{equation}
\label{eq:density}
n_e(r) = 10^7 D_A (1+z) \sqrt{4 \pi} \sqrt{\frac{1.2 \eta (r)}{V(r)}} ~ {\rm cm}^{-3}
\end{equation}
where $n_e$ is the electron number density, $D_A$ is the angular diameter distance at the cluster redshift $z$, $\eta$ is the {\scriptsize MEKAL} normalization, $V$ is the volume of the spherical shell, and the factor 1.2 comes from the ratio $n_e / n_p $ for a fully-ionized solar abundance plasma. The core density reaches $(2.8\pm0.8)\times 10^{-2}$ cm$^{-3}$ in the central bin, and peaks at $(4.8\pm0.7)\times 10^{-2}$ cm$^{-3}$ at R $\approx 10$ kpc. This density profile has a mostly continuous smooth slope, and was also fit with the analytic form from \citet{2006ApJ...640..691V}. The sudden jump in the last radial bin of the density and all derived radial profiles is an artefact from the deprojection process, which becomes less significant with smaller radii as the surface brightness is strongly centrally peaked in cool core clusters.

The total pressure profile was calculated using $P = 2 kT n_e$. The profile stays relatively smooth, reaching $(1.5\pm0.8)\times 10^{-10}$ erg cm$^{-3}$ in the central bin and peaking at $(2.9\pm1.1)\times 10^{-10}$ erg cm$^{-3}$ at R $\approx 18$ kpc. We rescale the universal pressure model from \citet{2007ApJ...668....1N} to the profile, and overlay it in \autoref{fig:rprof}. We will use this profile in \autoref{subsec:central_agn} to calculate the buoyant rise time and power in the radio bubbles.

\begin{figure*}
\centering
\includegraphics[width=\textwidth]{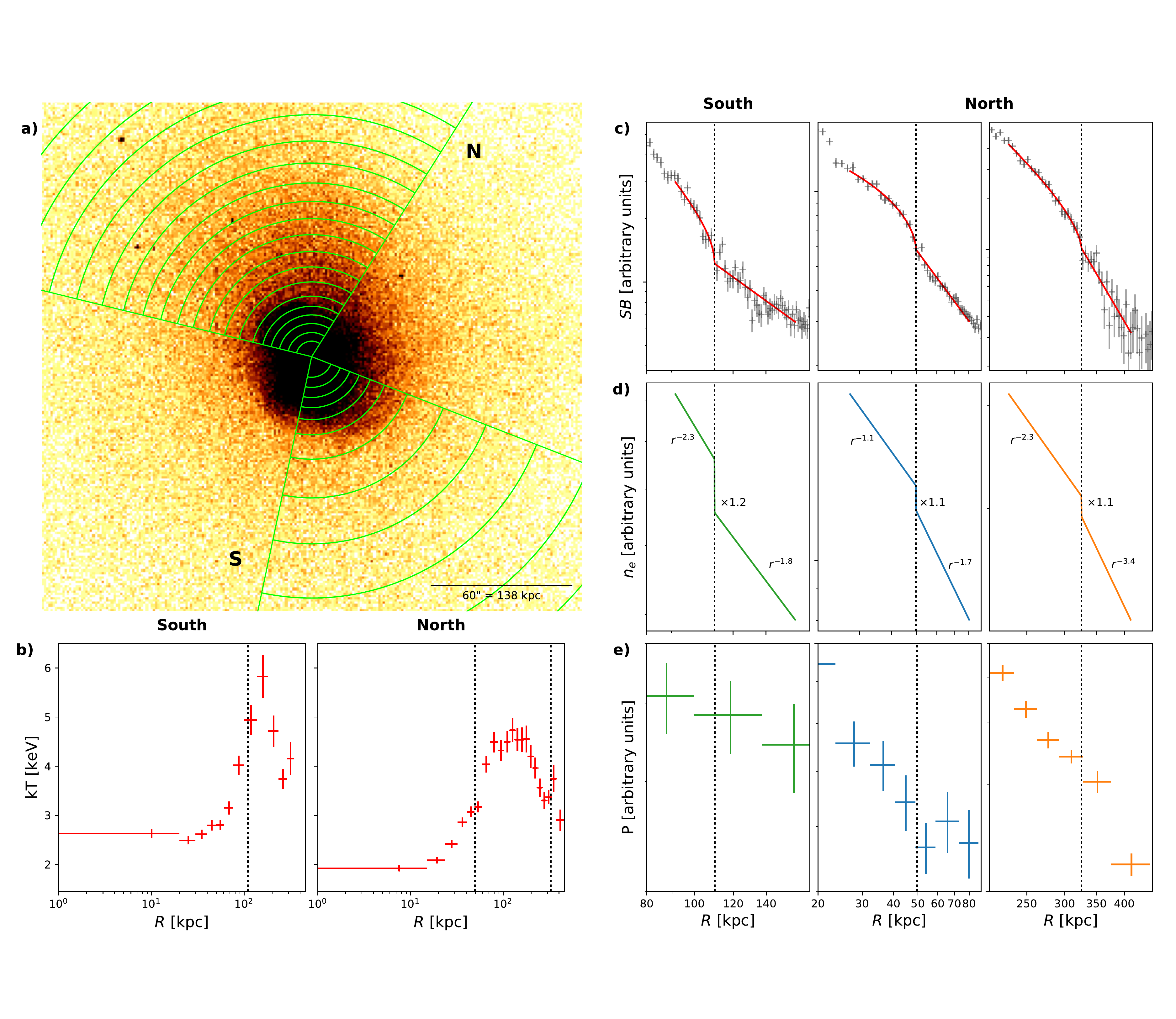}
\caption{Cold fronts in A1664. Panel \emph{\bf a)} X-ray image showing the regions used for extraction of (projected) temperature profiles, shown in panel \emph{\bf b)}. In panel \emph{\bf b}, the left-hand plots correspond to the (projected) temperature profile of the Southern sectors, while the right-hand plot corresponds to sectors in the Northern direction. In panels \emph{\bf c-e}, the left-most column corresponds to profiles in the Southern sectors, while the middle and right-most columns correspond to the Northern sectors depicted in panel \emph{\bf a)}, focusing on the inner and outer radii respectively. In all panels, the vertical axes have been scaled by an arbitrary factor to improve clarity. Panel \emph{\bf c)} 2D surface brightness profiles along the North and South directions, fit locally with the analytic line-of-sight projection of the 3D (i.e. deprojected) broken power-law density models given in panel \emph{\bf d)}. Panel \emph{\bf e)} shows the pseudo-pressure calculated from the product of panels \emph{\bf b)} and \emph{\bf d)}. Across each of the fitted density discontinuities, the pressure profile is smoothly varying, as is expected of a cold front in pressure equilibrium. \label{fig:rprof_hemi} }
\end{figure*}

The entropy profile was calculated using $S = kT n_e ^{-2/3}$. The profile falls below 30 keV cm$^2$ at a radius of 46 kpc (20\arcsec), a threshold below which \citet{2008ApJ...687..899R} find a higher occurrence rate of multiphase gas and ongoing star formation. Indeed, the ALMA observations of molecular gas in this system by \citet{2014ApJ...784...78R} support this scenario. Where this molecular gas might come from is discussed later in \autoref{sec:discuss}. The entropy profile reaches S = $19.1\pm9.1$ keV cm$^2$ in the central bin. We attempt to fit this profile with a power law of the form $S(r)\propto r^{1.2}$ following \citet{2005MNRAS.364..909V} for radii beyond $\sim 20$ kpc. \autoref{fig:rprof} shows that this single power law does not fit the data in the inner regions well, which are instead fit with a shallower power law, following the \citet{2018ApJ...862...39B} $S \propto r^{2/3}$ expectation and consistent with the results of, e.g., \citet{2014MNRAS.438.2341P}. The profile is mostly smooth, within errors, with the scatter in the data points from $R \approx 30-80$ kpc being artefacts from the deprojection of the temperature profile.

A cooling time profile was also produced using the equation 
\begin{equation}
\label{eq:tcool}
t_{cool} = \frac{5}{2} \frac{(n_e + n_I) k T}{n_e n_I \Lambda(T,Z)} = \frac{5}{2} \frac{n k T}{L / V} ~ ,
\end{equation}
where $n = n_e + n_I$ is the total number density of electrons and ions, respectively, $\Lambda(T,Z)$ is the metallicity and temperature-dependent cooling function to account for line cooling, $L$ is the unabsorbed X-ray model luminosity, calculated from running {\scriptsize CFLUX} on the thermal component in XSPEC (i.e.{\scriptsize WABS$\times$(CFLUX*MEKAL})), and $V$ is the volume of the corresponding spherical shell from which the spectrum is calculated. This flux was calculated over the range of 0.01--100.0 keV. The factor of 5/2 from the gas enthalpy in the cooling time calculation is used instead of the other frequently-used 3/2 factor (from thermal energy) since it is assumed that the plasma compresses as it cools, raising its heat capacity by 5/3 \citep{2006PhR...427....1P}. With this in mind, our cooling time profile is in agreement with \citet[hereafter K09]{2009ApJ...697..867K}, reaching a central cooling time of $(8.0\pm4.1)\times 10^{8}$ yr. We also overlay the ratio of the cooling time (based here on thermal energy rather than enthalpy) to the free fall time, $t_{ff} = \sqrt{2r/g}$, at each radius.

After correcting for the different assumed redshift values, N$_{\rm H}$ values and energy ranges used, our thermodynamic profiles agree with those found in K09, to within measurement uncertainties. We note, however, that the azimuthally averaged density profiles reported by K09 in Fig. 9 may actually be total number density, rather than electron density, related via $n = n_e + n_I \approx  1.91 n_e$. Upon comparison with the Archive of {\it Chandra} Cluster Entropy Profile Tables (ACCEPT) by \citet{2009ApJS..182...12C}, we find that our profiles agree with their published (projected) values as well, including the cooling time profile after accounting for the range (0.7 -- 2.0 keV) over which they calculated their cooling luminosities.

\subsection{Sloshing and Cold Fronts}
\label{subsec:sloshing}

Images of A1664 show that the surface brightness structure deviates from circular symmetry. As such, the spherical global profiles in \autoref{fig:rprof} smooth out any substructure along surface brightness edges. To explore potential deviations from symmetry, we extract spectra from circular wedges to the North and South of the cluster center, as illustrated in \autoref{fig:rprof_hemi}. These regions are spaced adaptively to enclose at least 5,000 counts. In addition, we sample the surface brightness profiles along these wedges in more finely-spaced radii. We expect that at the location of a surface brightness edge, the density should change abruptly. We model this discontinuity as a broken power-law with a discontinuous jump \citep{2007PhR...443....1M}, then project this three-dimensional model analytically (see \autoref{eq:analytic}) and locally fit to the observed, one-dimensional surface brightness profiles of the North and South sectors using a least-squares regression. The result of this fitting routine can be seen in panels \emph{c} and \emph{d} of \autoref{fig:rprof_hemi}. We stress that these fits are done locally, over a small region of the surface brightness profile (i.e. for $\sim 0.7 r_c$ to either side of each density discontinuity located at distinct radii $r_c$), where the assumption of a constant power-law slope is justified. In each of the panels \emph{b-e}, the left-hand plots correspond to the Southern sectors, while the right-hand plots correspond to sectors in the Northern direction. We achieve good fits to the surface brightness profiles, and find multiple edges, specifically at distances of about +55 kpc, -115 kpc, and +320 kpc ($+24$\arcsec, $-50$\arcsec, $+140$\arcsec)  from the X-ray centroid, where positive values refer to Northern radii, and negative values refer to edges located to the South. The factors by which the density jumps across each of these boundaries are 1.1, 1.2, and 1.1, respectively, which is characteristic of jump strengths across cold fronts.

At each of these radii, we investigate the (projected) temperature profiles as well (see panel \emph{b} in \autoref{fig:rprof_hemi}), and find that with each of the fitted density discontinuities, there is a corresponding jump in measured temperature. This behavior is expected across a cold front \citep[see][for a review]{2007PhR...443....1M}. Multiplying the model densities by the projected temperatures, we can calculate a pseudo-pressure, as seen in panel \emph{e} of \autoref{fig:rprof_hemi}. At the location of each edge, we find that the pressure is smoothly varying, as would be expected of a cold front in pressure equilibrium \citep{2007PhR...443....1M}.

\subsubsection{Temperature Map}

\begin{figure}
\centering
\includegraphics[width=\columnwidth,height=\columnwidth]{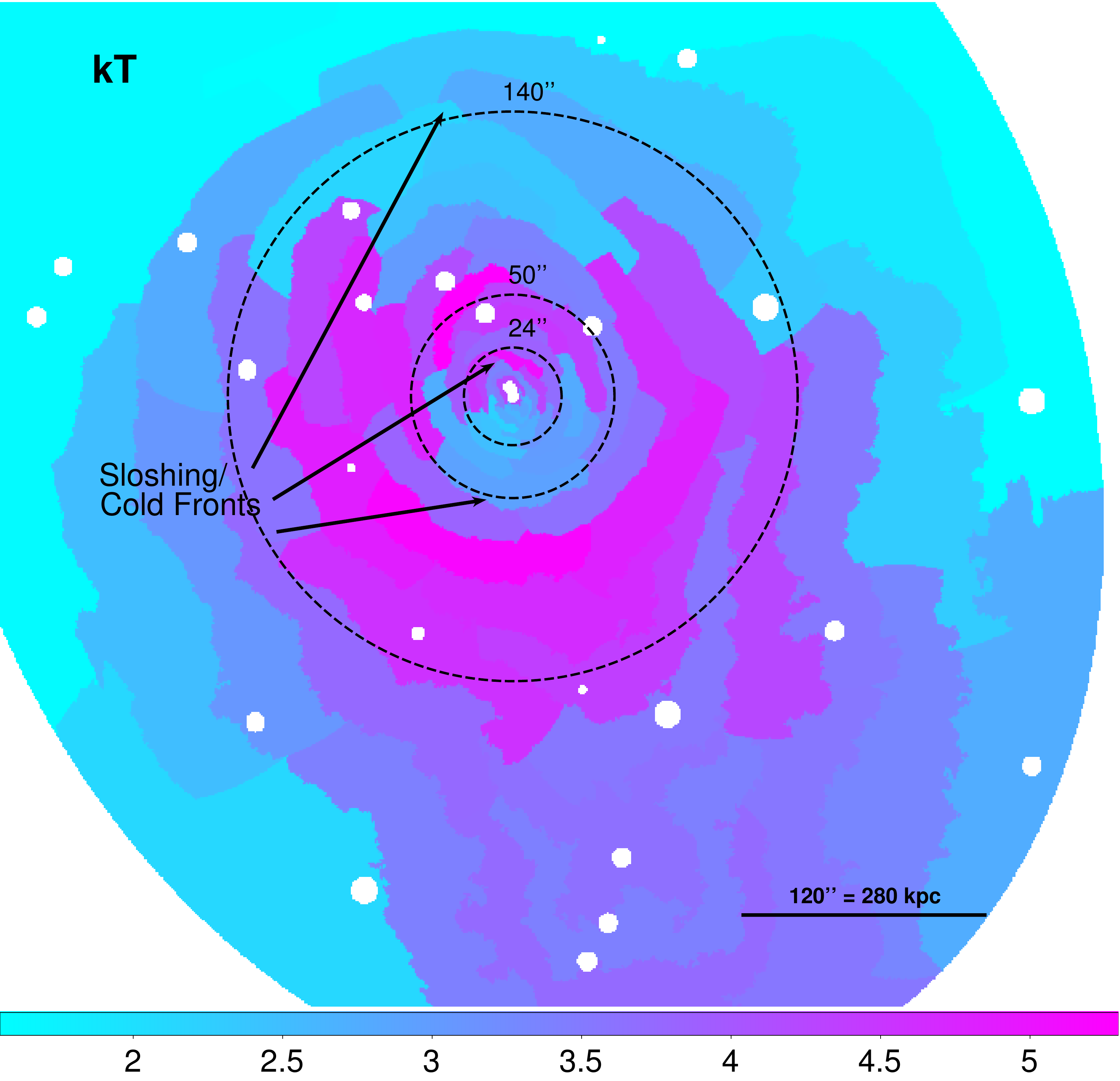}
\caption{Temperature map of A1664, in units of keV, produced via contour binning of the 0.5--7.0 keV events file image with pixels binned to 0.5\arcsec (bin=1) and approximately 5,000 counts per region (s/n = 70). Various edges in the NE-SW direction are visible, suggesting that the cool core may be oscillating in a disrupted gravitational potential . 
\label{fig:kT10000}}
\end{figure}

\begin{figure*}
\centering
\includegraphics[width=\textwidth,height=2.75in]{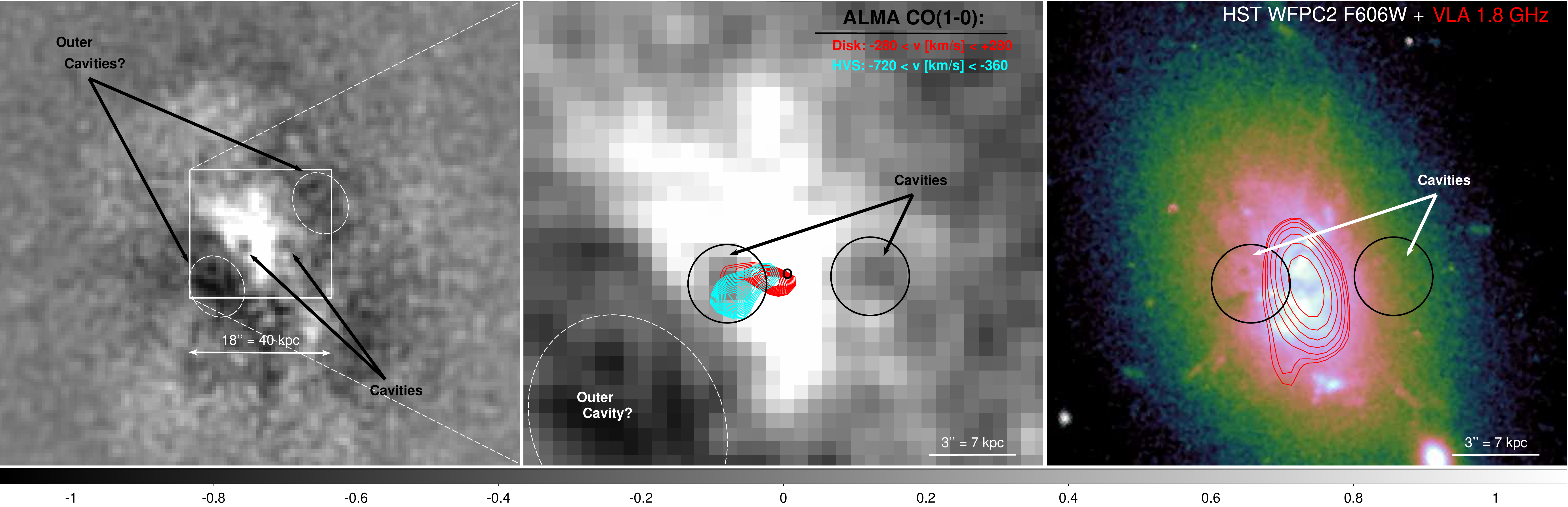}
\caption{\emph{Left}: {\it Chandra} residual flux image of A1664 after subtracting the sum of four freely-varying $\beta$-model fits, normalized by the flux image in the broad (0.5--7.0 keV) band, and binned such that one pixel is approximately 0.5\arcsec on a side. The image is smoothed with a 2-pixel radius Gaussian kernel. The image highlights a cavity system in the X-ray emission with a set of arrows toward the center of the cluster, and another set of possible outer cavities. \emph{Middle}: a zoomed-in view of the region of interest, smoothed with a 2-pixel radius Gaussian kernel. The `o' marks the location of the BCG. The contours from the molecular CO emission measured by {\it ALMA} are overlaid, and are found just behind or inside the putative Eastern cavity. The systemic possible "disk" component and the high-velocity system (HVS) are described in detail in \citet{2014ApJ...784...78R}. \emph{Right}: HST WFPC2 F606W image of A1664 for comparison, from \citet{2010ApJ...719.1619O}, with JVLA 1.8 GHz contours overlaid (courtesy of A. C. Edge). \label{fig:almaCO}}
\end{figure*}

To better study the spectral properties of A1664 in a way that follows its complex structure, especially in the inner core regions, we use the technique of `contour binning', devised by \citet{2006MNRAS.371..829S}, to create a temperature map. This algorithm bins X-ray data and creates regions for spectral extraction by following contours on an adaptively smoothed map of the image. Contour binning is especially good for clusters where the surface brightness distribution is asymmetric, and has been used already on a number of well-known clusters (see e.g. Centaurus: \citealt{2005MNRAS.360L..20F}; Perseus: \citealt{2006MNRAS.366..417F,2005MNRAS.360..133S}).

This method was used to produce a (projected) temperature map of the core of A1664, as seen in \autoref{fig:kT10000}. These regions were chosen to contain approximately 5,000 counts. One can immediately see an elongation of structure in the NE-SW direction, with warm (3.5 -- 4.0 keV) gas trailing Southward. The ICM core seems to be sloshing back and forth in this direction, where it first created the cold front located now at 320 kpc (140\arcsec) North from the center, then turned around and created another cold front which is now found at a distance of 115 kpc (50\arcsec) to the South, and turned around once more creating the cold front at 55 kpc (24\arcsec) to the North of the X-ray centroid. These distinct cold front sites `radiate' outward from the center of a cluster in a slow, continuous wave as a result from gas sloshing perturbations. These perturbations are composed mainly of dipolar g-modes that are excited in a merger. Initially, these modes are co-aligned, but they rotate at about the Brunt-Vaisala (BV) frequency as they evolve, forming a spiral pattern that wraps outward from the center of a cluster as the BV frequency decreases with increasing radius \citep[see e.g.][]{2011MNRAS.413.2057R}. The previous analysis of K09 revealed a very clear residual spiral structure in A1664 ( on larger scales than that shown in \autoref{fig:almaCO}) after subtracting from its emission a beta model. Also, we see that the region to the south is hotter and denser than the rest of the cluster, implying that it is overpressured. This suggests that the merger is quite recent, not more than about a sound crossing time ($t_c = \frac{320 ~{\rm kpc} }{600 ~{\rm km ~s}^{-1}}  \approx 0.5$ Gyr) in the past. This is a rough estimate, however, and better constraints on this timescale may be provided by the location and age of the radio relic to the South. \citet{2014MNRAS.439.2755K} used {\it BRI} photometry and shallower \emph{Chandra} observations to point out substructure $\sim800$ kpc South of the cluster core, and identify it to most likely be the remnant core of a merging group which has passed pericenter. This possible merger may be the one responsible for triggering the cold front we observed here, and future observations of this substructure may yield further constraints on the time since last merger.

It is also worth noting that the 1.8 GHz JVLA data revealed a wide-angle tailed (WAT) radio source $\sim 600$ kpc $(260\arcsec)$ to the South of the cluster core, which is another indicator that the ICM is disturbed. In addition, in a far ultra-violet (FUV) study of 16 low-redshift ($z<0.3$) cool core BCGs by \citet{2015MNRAS.451.3768T}, A1664 had among the most disturbed FUV morphology, as measured by anisotropy index. A1664's highly-disturbed FUV morphology is further evidence that there is core sloshing, affecting even lower temperature material. Also, the Ly$\alpha$ morphology shows a filament extending $\sim 15$ kpc Southward from the BCG, in the same direction as the sloshing.

\subsection{Radiative and Mechanical Properties of the Central AGN}

\subsubsection{Luminosity of Nuclear Point Source}
\label{subsubsec:l_nuc}

There is a point source visible at the cluster center, which is detected in the hard (2.0--7.0 keV) X-ray band at the 3.3$\sigma$ level. The coordinates of the brightest pixel, taken to be the nucleus, are $\alpha = 13^h03^m42 \fs 592$, $\delta = -24^{\circ}14'42 \farcs 89$, about 5.3 kpc $= 2.3\arcsec$  from the X-ray centroid found above, and marked by a `x' in \autoref{fig:hardcavity}. The X-ray point source is apparently $\sim 0.7\arcsec$ offset from the Very Long Baseline Array (VLBA) 5 GHz position of (RA, Dec) = ($13^h03^m42\fs565, -24^{\circ}14'42\farcs218$ \citep{2014PhDT.......338H}, though this offset is likely not significant given the possible astrometric errors between \emph{Chandra} and the radio reference frame. \citet{2012MNRAS.424..224H} gave an upper limit on the luminosity of this point source of ${\rm log}~ L_X < 41.18 \pm 0.48$ over the 2--10 keV band. We utilize the same method, and find a nuclear luminosity of ${\rm log}~ ( L_X /{\rm [erg ~s^{-1}]}) = 40.93 \pm 0.04$.

Following the methodology of \citet{2011MNRAS.413..313H}, we also calculate an upper limit on the spectroscopic	luminosity of the nuclear point source by extracting a spectrum from a circular region around the brightest X-ray pixel identified in the hard band (3--7 keV) with a radius of 1.5\arcsec (representing about 95\% of the encircled energy for \emph{Chandra's} on-axis PSF\footnote{\url{http://cxc.harvard.edu/proposer/POG/html/chap4.html}}), which contains $\sim$1,269 counts in the 0.5--7.0 keV band. We fit this spectrum over the 0.5--7.0 keV band and model the AGN emission with XSPEC model {\scriptsize POWERLAW}, and the thermal emission with {\scriptsize MEKAL} while accounting for absorption. We calculate the unabsorbed AGN luminosity with the {\scriptsize CFLUX} model in the following way: {\scriptsize (WABS$\times$(MEKAL+CFLUX*POWERLAW))}. We perform the fit with a fixed column density, power law spectral index of $\Gamma = 1.9$, and fix the temperature and abundance to 2.1 keV and 0.39 Z$_{\odot}$, which were the values of the innermost bin from our projected. The resulting unabsorbed AGN model luminosity in the 2--10 keV band is only an upper limit, with $L_X < 2.3\times 10^{41}$ erg s$^{-1}$, in agreement with \citet{2012MNRAS.424..224H}. This spectroscopic method has greater degrees of freedom than the photometric method, so the constraints on the AGN's luminosity are weaker.

\subsubsection{The X-Ray Cavities}

\begin{figure}
\centering
\includegraphics[width=\columnwidth,height=3in]{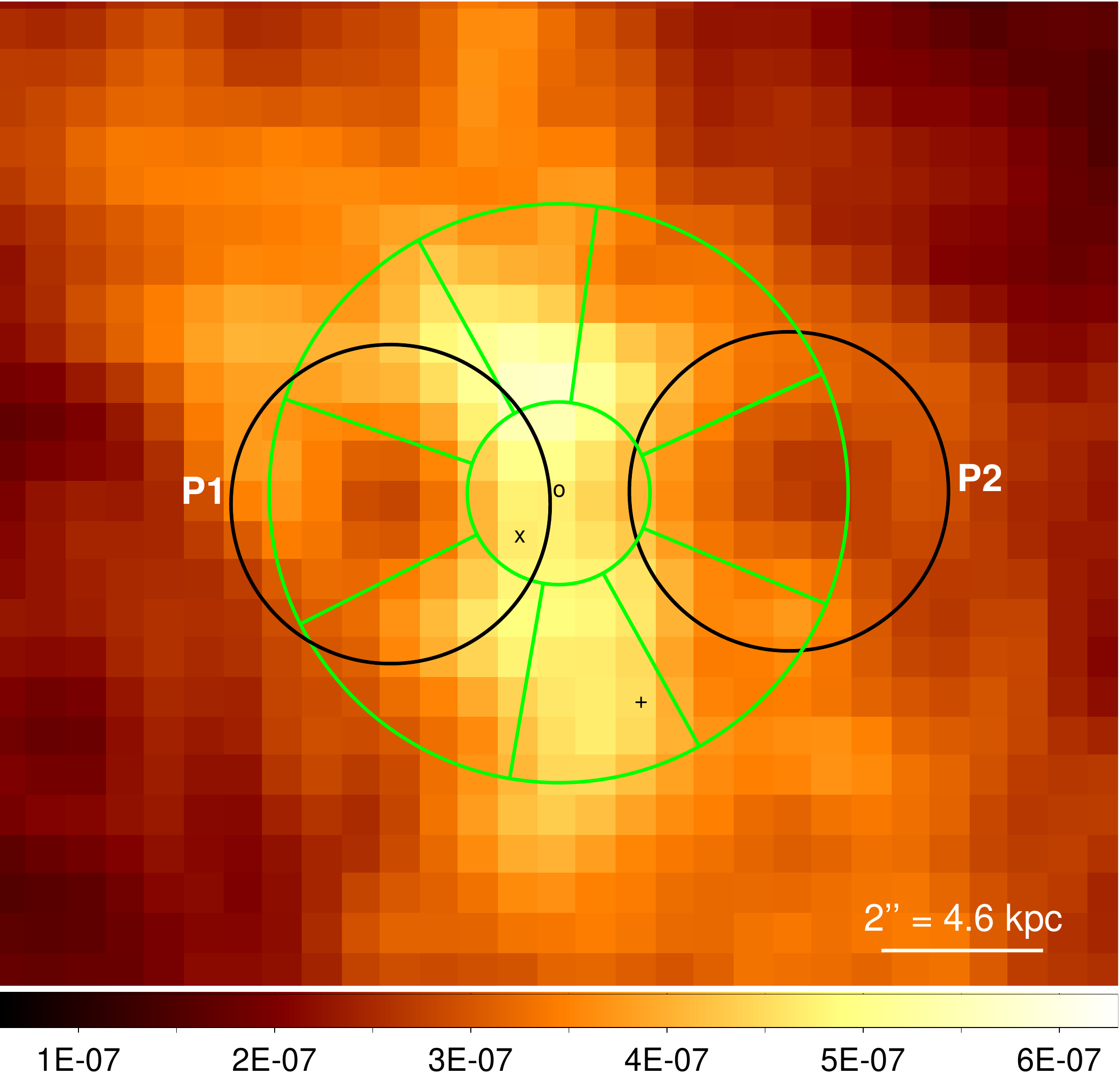}
\includegraphics[width=\columnwidth,height=3in]{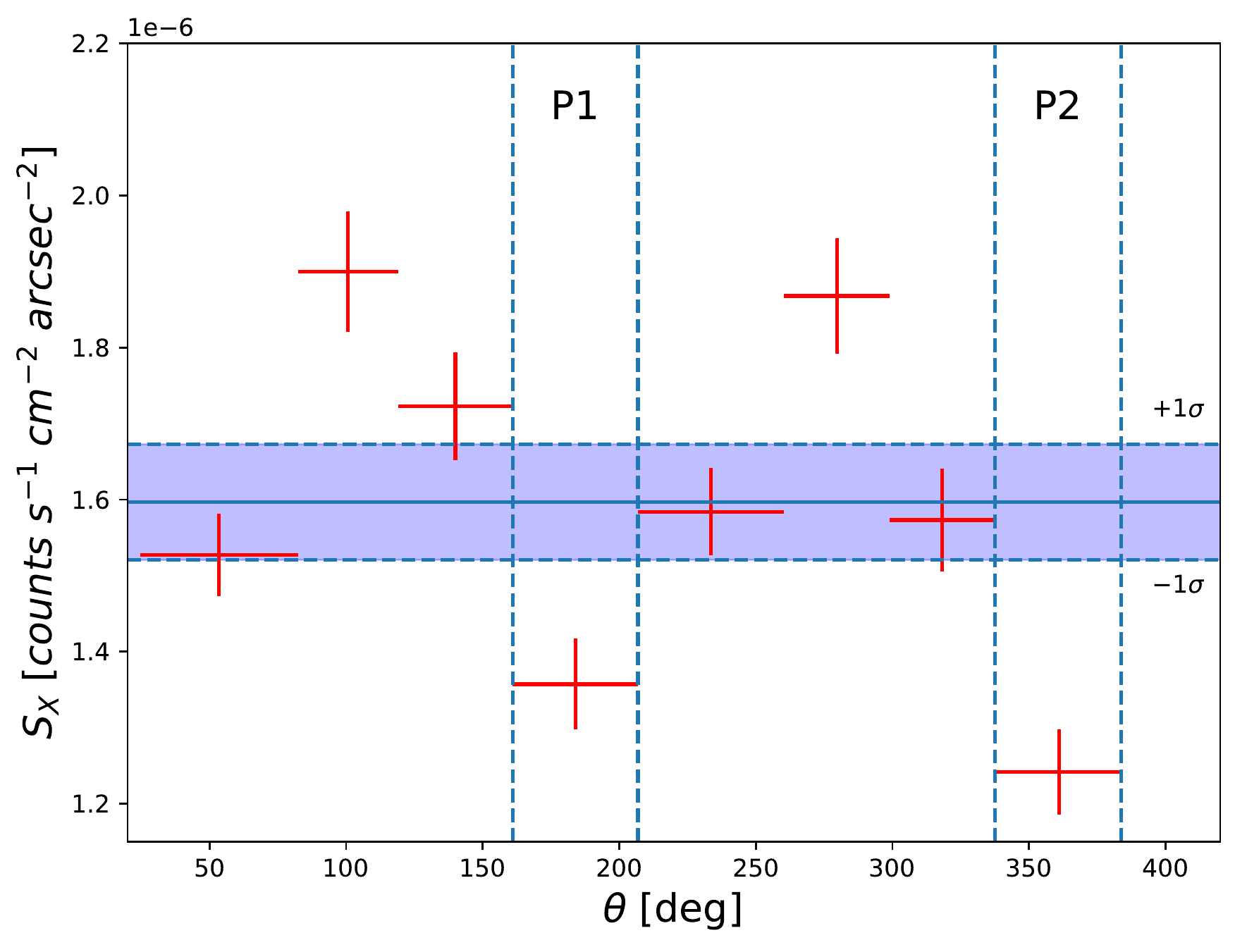}
\caption{\emph{Top}: 0.5--7 keV broadband exposure-corrected flux image, binned such that 1 pixel is approximately 0.5\arcsec on a side, and smoothed with a Gaussian with a width of 2\arcsec. The sectors highlight the azimuthal surface brightness extraction regions, with the two possible cavities P1 and P2 identified from \autoref{fig:almaCO}. Note that the cavities are not centered on the BCG, denoted on this image with a black circle. The `+' symbol marks the X-ray centroid found from the beta-model fitting, while the `x' symbol marks the location of the brightest pixel in the 3--7 keV hard X-ray background. \emph{Bottom}: Background-subtracted azimuthal surface brightness measurements of the sectors along the expected cavities relative to adjacent sectors for comparison, centered on the BCG. The angles for each sector are measured counter-clockwise from `W' on the sky. The statistical significance of both of the potential cavities P1 and P2 is $\sim4.0\sigma$ and $6.4\sigma$, respectively. Note that the two statistically significant positive increments correspond to the North-South X-ray bar structure. \label{fig:hardcavity}}
\end{figure}

\citet{2014MNRAS.444.1236P} show that $\gtrsim$ 20,000 counts are required in the central 20 kpc of a cool core cluster to detect the presence of X-ray cavities. A1664 has approximately 22,000 counts in the 0.5--7.0 keV band within the central 20 kpc, which ought to be sufficient for a strong detection. As such, we proceed with a search for these cavities, knowing that the data are of sufficient depth and quality.

\autoref{fig:almaCO} illustrates the possible existence of two pairs of cavities where there are depressions in X-ray surface brightness to the East and West of the BCG. \autoref{fig:almaCO} is a residual image created by fitting four beta-models to the flux image, as in \autoref{sec:results}, then subtracting the combined fits from the original flux image and normalizing by it. The more tentative ``outer cavities'' are marked by circular dashed regions, and are located roughly 23 kpc (10\arcsec) away from the BCG. The potential inner cavities are oriented perpendicular to the X-ray bar, whose presence may be the result of sloshing. The third panel in \autoref{fig:almaCO} displays the HST WFPC2 F606W image of A1664 for the same field of view as that in the middle panel. The central galaxy looks highly disturbed, with filamentary structure in several directions as well as dust lanes co-spatial with the ALMA contours. Overlaid on the HST image are 1.8 GHz contours of the radio source detected with the JVLA, provided courtesy of A. C. Edge. There is no evidence of radio emission associated with the inner cavities, but it is unresolved at this frequency, and many other BCGs do not have observed radio lobes. In addition, the lack of radio emission at the location of the cavities may be due to spectral aging. Indeed, A1664 has a steep spectral index below 1 GHz, so 300 MHz observations at 1$\arcsec$ resolution, for instance, should reveal definitively whether there is radio emission at the locations of the cavities.

To determine the significance of these potential inner cavities, we compare the surface brightness in these regions with that of adjacent regions, as shown \autoref{fig:hardcavity}. We have divided the area of interest into eight annular wedges centered on the BCG, with inner and outer radii of $1.1\arcsec$ and $3.6\arcsec$, respectively. The two black circular regions are overlaid on the location of the potential inner cavities, with the potential East cavity sector labeled `P1', and the West cavity labeled `P2'. In the panel below is a plot of the azimuthal surface brightness measurements as a function of the range of angles spanned by each sector. The average surface brightness, estimated over all eight regions assuming the null hypothesis that the depressions in surface brightness are noise, is $S_X = (1.60 \pm 0.08) \times 10^{-6}$ counts s$^{-1}$ cm$^{-2}$ arcsec$^{-2}$. The error on this average surface brightness was calculated by taking the standard deviation of all eight surface brightness measurements then dividing by the square root of the number of measurements (i.e. 8). The surface brightness of P1 (P2) is $(1.36\pm0.06) \times 10^{-6} ~ ((1.24\pm0.06) \times 10^{-6})$ counts s$^{-1}$ cm$^{-2}$ arcsec$^{-2}$, making these inner cavities $4.0\sigma$ and $6.4\sigma$ detections respectively, and we thus reject the null hypothesis. We note two caveats to these conservative detection significance levels. First, our regions were chosen to increase the significance of these cavities, which will artificially raise the significance of this detection. On the other hand, the fact that there are two statistically significant decrements that are diametrically opposed (180$^{\circ}$ offset) further strengthens the overall detection. We will discuss the implications of these confirmed inner cavities and the potential outer cavities in \autoref{subsec:central_agn} and \autoref{subsec:alma_origin}.


\section{Discussion}
\label{sec:discuss}

\subsection{AGN Feedback in Abell 1664}
\label{subsec:central_agn}

In the following discussion, we derive a cavity power as a proxy for AGN power. The detected inner cavities are at a distance of $d \sim 4.6\pm1.1$ kpc ($2.0\pm 0.5$\arcsec) and $d \sim 6.7\pm1.1$ kpc ($2.9\pm 0.5$\arcsec) from the BCG position, for the East and West cavities respectively, with approximately equal radii of $r \sim 4.6 \pm 1.1$ kpc ($2.00\pm 0.5$\arcsec). The distance and size uncertainties here are based on the spatial resolution of \emph{Chandra}. These do not, however, reflect our uncertainty on the line of sight extent of each cavity. 

We proceed to calculate the AGN power via the cavity power, $P_{cav}$ as follows:
\begin{equation}
P_{cav} = \frac{4 p V}{t_{\rm buoy}},
\end{equation}
where $4pV$ is the enthalpy of a cavity filled with relativistic fluid, and $t_{\rm buoy}$ = $d \sqrt{S C_D/2 g V}$ is the age of the cavity based on the buoyancy rise time, where $d$ is the projected distance of the cavity from the cluster core, $S=\pi r^2$ is the bubble's cross section, $C_D=0.75$ is the drag coefficient \citep[e.g.,][]{2001ApJ...554..261C}, the volume $V$ = $\frac{4}{3} \pi r^3$, where $r$ is the cavity radius, and $g \simeq 2 \sigma^2 /d$ is the local gravitational acceleration. $g$ can be estimated using the local stellar velocity dispersion $\sigma$, which was measured by \citet{2018ApJ...853..177P} to be $267\pm12$ km s$^{-1}$ for A1664.
One can also calculate the gravitational acceleration via a mass estimate, but we note that the pressure profile we measure is poorly resolved on the scale of interest. Using the above values for the East cavity, we find a buoyancy rise time of $t_{\rm buoy} = (6.3\pm2.4)\times 10^{6}$ yr, but we note that this age is likely underestimated due to projection effects. As a consequence, after multiplying by a factor of two for the West cavity, we calculate a cavity power of $P_{cav} = (1.1 \pm 1.0) \times 10^{44} $ erg s$^{-1}$. This cavity power value agrees, within the uncertainties, with the value obtained by K09, who made their calculations using scaling relations between jet mechanical power and radio synchrotron power \citep{2008ApJ...686..859B}, as they did not directly detect cavities in their shallower X-ray data. 

The inner regions of this cluster have a short cooling time, although only a small fraction of the total gas is cooling. To determine a cooling luminosity, we took a spectrum of the cooling region, which we define to be where the cooling time falls below $\sim 3 \times 10^9$ yr \citep[e.g.][]{2018ApJ...858...45M}. For A1664, this cooling region corresponds to the inner $\sim$70 kpc (30\arcsec) (see \autoref{fig:rprof}). This region was fitted with a single-temperature {\scriptsize MEKAL} model with fixed column density. These fits yielded a bolometric (0.01--100.0 keV) luminosity of $(1.90 \pm 0.01) \times 10^{44}$ erg s$^{-1}$. This luminosity is comparable to the AGN jet power estimated from the cavities, of $P_{cav} \sim (1.1 \pm 1.0) \times 10^{44} $ erg s$^{-1}$. Thus, given the uncertainty in estimating the size of the cavities, coupled with the fact that cavity powers tend to underestimate mean jet powers by a significant factor \citep{2007ARA&A..45..117M} and that the outburst powers vary with time, the AGN in A1664 may also be powerful enough to offset the cooling of the ICM. 

We also determined an upper limit for the nuclear X-ray luminosity of A1664 to be $\lesssim $ 2.3$\times ~ 10^{41}$ erg s$^{-1}$ (see \autoref{subsubsec:l_nuc}). This is orders of magnitude smaller than the cooling luminosity of $1.90 \times 10^{44}$ erg s$^{-1}$, and the inferred jet power from the possible cavities of $P_{cav} \sim 1.1 \times 10^{44} $ erg s$^{-1}$, which is consistent with the picture that this AGN is radiatively inefficient and prevents cooling mechanically via outflows, bubbles, or sound waves rather than through radiation, similar to many other BCGs in the nearby universe \citep[see, e.g.,][]{2013MNRAS.432..530R}.

\begin{figure*}
\centering
\includegraphics[width=\textwidth, height=2.0in]{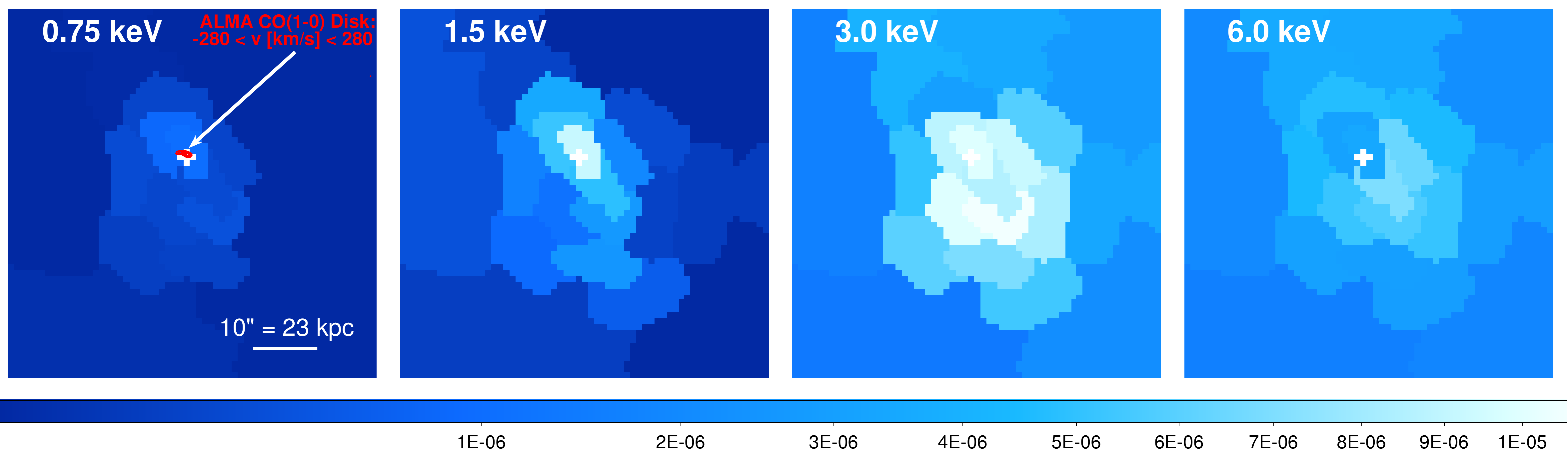}
\caption{Multi-temperature {\scriptsize MEKAL} maps showing the XSPEC normalization per unit area of corresponding regions, in units of cm$^{-5}$ arcsec$^{-2}$, where the regions were made via contour binning to secure at least $\sim$5,000 counts per region. The central point source has been masked. At high temperatures there is a screen of material due to projection of gas shells on the line of sight. At lower temperatures, most of the cool gas lies along the X-ray bar, and is especially concentrated to the North of the cluster center, possibly cooling towards the molecular gas flowed centered on the BCG's systemic velocity seen with {\it ALMA} (red contours). All panels are on the same spatial and colorbar scale. \label{fig:multiapec}}
\end{figure*}

\subsection[Origin of the Molecular Gas]{The Origin of the Cold Molecular Gas}
\label{subsec:alma_origin}

As mentioned before, at the core of A1664 lies a reservoir of cold molecular gas, first detected by \citet{2001MNRAS.328..762E}, as well as a complex distribution of disturbed molecular hydrogen \citep{2009MNRAS.395.1355W}. The molecular gas was later observed with {\it ALMA} by \citet[hereafter R14]{2014ApJ...784...78R}, revealing $\sim 10^{10} ~ {\rm M}_{\odot}$ of molecular gas distributed evenly over two distinct velocity systems: a molecular gas reservoir centered on the BCG's systemic velocity, which could be fueling accretion onto the SMBH on smaller scales, and a high-velocity system (HVS) indicating a gas flow at 600 km s$^{-1}$, at a distance $\sim$ 11 kpc (5\arcsec) from the nucleus. It is hard to determine without further observations of absorption lines whether the blueshifted velocities in the HVS are due to an inflow of gas, if behind the BCG, or an outflow positioned in front of the BCG along the line of sight.

If the HVS is an outflow, it is likely that the AGN is driving that gas flow, perhaps by uplift from buoyantly-rising radio bubbles \citep[e.g.][]{2010MNRAS.406.2023P}. In a similar study, \citet{2014ApJ...785...44M} reveal a $10^{10} ~ M_{\odot}$ high-velocity molecular gas system in A1835 that extends along a low-surface brightness channel in the X-ray emission and towards two cavities located on either side of the nucleus. ALMA observations have shown that massive molecular gas filaments extend toward and are drawn up around cavities in other targets, PKS 0745-191 \citep{2016MNRAS.458.3134R}, A2597 \citep{2016Natur.534..218T}, A1795 \citep{2017MNRAS.472.4024R}, 2A 0335+096 \citep{2016ApJ...832..148V}, and Phoenix \citep{2017ApJ...836..130R}. We have shown that A1664 harbors such a cavity system, and we postulate that the HVS is an outflow potentially driven by the central AGN. The cavity power of $\sim (1.1 \pm 1.0) \times 10^{44} $ erg s$^{-1}$ given previously in \autoref{subsec:central_agn} is comparable to the kinetic power of the molecular gas outflow of $\sim 3 \times 10^{43}$ erg s$^{-1}$, calculated by R14. Therefore given the uncertainty on the cavity size and the fact that cavity powers tend to underestimate jet powers, this jet power is likely  sufficient to drive the molecular gas outflow after accounting for various sources of energy loss.

The East cavity `P1' is closer to the BCG center (in projection) than the West cavity `P2'. This may be exaggerated by projection effects. R14 indicated that the HVS observed with {\it ALMA} has a broad velocity shear, implying that the acceleration of these gas clumps may be almost parallel to the line of sight. Thus, it is plausible that if this HVS is an outflow, that the East bubble dragging out the gas is also rising along the line of sight. By Archimedes' principle, these bubbles could not lift up more mass than they displace. The electron number density at the location of the bubbles from the X-ray centroid ($d \sim 2$ arcsec) is $n_e \approx $ 0.05 cm $^{-3}$, so for a single bubble with a radius of $\approx$ 4.6 kpc (2\arcsec), the average mass displaced by one would be $m = \rho_{\rm gas} V = \mu m_p n \frac{4}{3} \pi (4.6 {\rm kpc})^3 = 1.2 \times 10^{42}$ g, or $5.9 \times 10^8$ M$_{\odot}$, where $\mu = 0.61$ is the mean molecular weight. It is thus not possible to create the HVS with a mass of $\sim 5 \times 10^9$ M$_{\odot}$ entirely via direct uplift by the bubble. There is some evidence (see \autoref{fig:almaCO}) of potential ``outer cavities'' further from the cluster center, so it is a possibility that the molecular gas cooled over multiple AGN outburst cycles \citep[e.g., as in A1795:][]{2017MNRAS.472.4024R}. Thus, rather than direct uplift, it is more likely that the molecular gas is cooling in situ behind the bubble, as the bubble lifts up warmer gas, increasing its infall time and promoting condensation into molecular clouds \citep[see e.g.][]{2016ApJ...830...79M}. Nevertheless, it is worth noting that the molecular gas mass calculation is critically dependent on the correct calibration of the CO-to-H$_2$ conversion factor $X_{\rm CO}$, which depends critically on environmental factors, and may be up to a factor of $\sim 2 \times$ lower than Galactic in BCGs \citep{2017ApJ...848..101V}, though this would not help the factor of $\sim 10 \times$ discrepancy in the displaced versus molecular gas masses.

\bigskip

\subsubsection{Tracing the Multi-phase Gas}

On the other side of this `cold phase' of AGN feedback is the matter of fueling the AGN through cold accretion of gas that condenses out of the cluster's hot atmosphere (e.g. \citealt{2010MNRAS.408..961P}). A1664's molecular gas reservoir centered on the BCGs systemic velocity from R14 is potentially fueling accretion onto the SMBH on small scales \citep[see also][]{2009MNRAS.395.1355W}. This gas flow could be a nascent disk, but it would clearly be unsettled, with little indication of ordered motion and a lopsided mass distribution. Thus, it is necessary to trace the most rapidly cooling X-ray gas component that is likely feeding the possible inflows, and thus show how the cluster atmosphere on tens of kpc scales may be linked to the central AGN. 

\autoref{fig:multiapec} shows the amount of gas at different temperatures for regions chosen via the contour binning method to have $\gtrsim$5,000 counts (signal-to-noise = 70), with the central point source excluded. Here we fit multiple fixed-temperature {\scriptsize MEKAL} components (at kT=0.75, 1.5, 3.0, and 6.0 keV), fixing the column density and fixing the abundance parameter of each component to $Z=0.65 Z_{\odot}$ (found from fitting the total spectrum of the inner $\sim$70 kpc (30\arcsec) with the same model and allowing abundance to vary, see \autoref{fig:rprof}), while allowing the normalizations to vary, following the technique described in \citealt{2006MNRAS.366..417F} (see also \citealt{2016MNRAS.457...82S}). At 6.0 keV, we essentially see a screen of hotter gas in projection, but concentrated especially in the X-ray bar at 3.0 keV. At 1.5 keV, the gas is still concentrated along the bar, but appears to be most abundant slightly North of the core, possibly towards the molecular gas reservoir centered on the BCG from \citet{2014ApJ...784...78R}. This region is slightly enhanced still in the 0.75 keV map, albeit more faintly as these maps have been normalized to the same intensity scale. In the 0.75 keV map, the innermost region has a normalization per unit area of $(1.0\pm0.3) \times 10^{-6}$ cm$^{-5}$ arcsec$^{-2}$, while the next brightest region, just South of the core, is $(3.4\pm2.0) \times 10^{-7}$ cm$^{-5}$ arcsec$^{-2}$.

In hot cluster atmospheres, the local cooling time of the ICM appears to correlate with the presence of thermal instabilities, such that we observe multiphase gas only where the cooling time of the ICM drops below ~1 Gyr \citep{2008ApJ...683L.107C,2008ApJ...687..899R,2010ApJ...721.1262M}. More recent work \citep{2012ApJ...746...94G,2012MNRAS.419.3319M,2012MNRAS.420.3174S,2017MNRAS.466..677G,2015Natur.519..203V,2016ApJ...830...79M} has suggested that there is an additional timescale that is important, akin to a mixing time, and that the ratio of the cooling time to the mixing time is the relevant quantity for predicting whether multiphase gas will condense from the cooling ICM. We find, in agreement with \citet{2018ApJ...853..177P}, that the ratio of the cooling time to the freefall time for A1664 is in the range of 10--20 in the inner $\sim$50 kpc (see \autoref{fig:rprof}), consistent with the picture of precipitation-regulated feedback \citep{2017ApJ...845...80V}.  The ALMA observations of a dozen systems show that the coldest gas lies predominantly in filaments that are projected behind radio bubbles, similar to the HVS in A1664.  \citet{2016ApJ...830...79M} \citep[see also][]{2017ApJ...845...80V} suggest that uplift by radio bubbles is promoting thermally unstable cooling and that the infall timescale may dictate the formation of these cold filaments. The specific details of how thermal instabilities develop are beyond the scope of this paper, but we can address whether there is sufficient cooling to fuel the ongoing star formation and massive cold gas reservoir in the central galaxy.

To determine whether it is likely that cooling instabilities and inflow could be feeding the observed molecular gas reservoirs, we investigate the amount of X-ray gas that can cool by fitting a {\scriptsize WABS$\times$(MEKAL+MKCFLOW}) model to the 30\arcsec ~region around the cluster core, with the upper temperature of the cooling flow tied to the temperature of the thermal component, and the lower temperature fixed at kT = 0.1 keV to represent full cooling (FC), or past the detectable range for {\it Chandra} \citep[e.g.][]{2004ApJ...601..184W}. We obtain a mass deposition rate of 42 $\pm$ 3 $M_{\odot}$ yr$^{-1}$, in agreement with K09, within errors. This deposition rate is a factor of a few higher than the published star formation rate of 13 $\pm$ 1 $M_{\odot}$ yr$^{-1}$ \citep{2018ApJ...858...45M}. Given that there are $\sim 10^{10} ~ M_{\odot}$ of molecular gas, the mass deposition rate quoted here also means this system would take $\sim 2 \times 10^{8}$ yr to form enough molecular gas. Such a timescale is a factor of a few lower relative to similar systems, and longer than the buoyancy timescale for the cavities of $t_{\rm buoy} = (6.3\pm2.4)\times 10^{6}$ yr. Non-radiative cooling may be playing a role here, where the hot gas interpenetrates the cold molecular gas \citep[e.g.][]{2011MNRAS.417..172F}, or the molecular gas could have formed over multiple AGN feedback cycles, as suggested by the potential ``outer cavities'' seen in \autoref{fig:almaCO}.

Since the X-ray mass deposition rate is consistent with sufficient gas cooling out, the multi-phase gas structure presented here in \autoref{fig:multiapec} is consistent with the hypothesis that the molecular gas clouds observed by R14 formed in-situ via cooling instabilities \citep{2014MNRAS.439.2291W}. This would then favor the scenario in which the gas reservoir centered on the BCG's systemic velocity is actually an inflow, which is supported by its highly asymmetric mass and velocity structures (see R14). The fact that the coolest X-ray gas is spatially coincident with the cold CO gas suggests that it may be fueling the cooling molecular gas, which has been seen in many other systems \citep[e.g.][]{2003A&A...412..657S,2006A&A...454..437S,2016Natur.534..218T,2017MNRAS.472.4024R}.

\section{Summary \& Conclusions}
\label{sec:end}

We have presented here new {\it Chandra} X-ray data, which revealed rich structure, including some elongation and accompanying compressions of the X-ray isophotes in the NE-SW direction, indicative of a gas core sloshing in the gravitational potential. This motion has resulted in cold fronts, which have expanded outwards and are now located at distances of about 55, 115, and 320 kpc (24, 50 and 140\arcsec) from the cluster center. These cold fronts are confirmed by looking at a detailed temperature map of A1664, where there are regions of high contrast across the edges. 

We conclude that the core of A1664 is highly disturbed, as the global metallicity and cooling times flatten at small radii, implying mixing of low metallicity gas. We found that A1664 hosts a faint X-ray point source at its center, and were able to determine its luminosity photometrically, and an upper limit spectroscopically. The radiative output of this AGN is orders of magnitude smaller than its mechanical power output and ICM cooling luminosity, implying that its black hole may be extremely massive and/or radiatively inefficient.

We found that the AGN has also undergone a mechanical outburst, as can be seen from our detection of inner cavities to the East and West of the BCG, with buoyant rise times of about $t_{\rm buoy} = (6.3\pm2.4)\times 10^{6}$ yr. This cavity system represents a total power of $P_{cav} = (1.1 \pm 1.0) \times 10^{44} $ erg s$^{-1}$, which is comparable to the cooling luminosity of 1.90 $\times 10^{44}$ erg s$^{-1}$. These X-ray cavities are the result of radio bubbles inflated by the AGN jet, which may be able to partly explain the presence of the massive molecular gas flows present near A1664's BCG, previously detected with {\it ALMA}. These data reveal roughly 10$^{10}$ M$_{\odot}$ of molecular gas. Roughly half of this cold gas is in a molecular gas reservoir with smooth velocity structure centered in velocity space on the BCG's systemic velocity, and centered spatially on the coolest X-ray emitting gas.

The remaining cold molecular gas is a high-velocity system (HVS) at 600 km s$^{-1}$ with respect to the BCG's systemic velocity. If this HVS is an outflow, it is possible that it is being drawn out from the cluster center by the AGN bubbles, as no other process is energetically feasible. In this and other systems, there exists a spatial coincidence between cavities and molecular gas. It is possible that the bubbles inflated by AGN are energetically capable of pulling up colder molecular gas in their wake as they rise buoyantly through the ICM, but it is still unclear whether the molecular gas gets drawn out directly, cools in situ, or perhaps even falls back in around the cavities. In the case of A1664, there is sufficient energy to lift the cold gas, but the amount of mass displaced by the bubbles is less than that of the molecular gas flows, so it is not possible for them to directly uplift the molecular gas, by Archimedes' principle. However, there is some evidence of potential ``outer cavities'' in this system, which would indicate multiple outbursts. The presence of older cavities would make it more likely that, rather than via direct uplift, the molecular gas is cooling in situ behind the bubble, as the bubble lifts up warmer gas, increasing its infall time and promoting condensation into molecular clouds. We use the extent and age of A1664's bubbles/cavities to calculate the mechanical jet power of the central AGN, and determine that this cavity power could be energetic enough to prevent the bulk of the ICM from cooling.

\acknowledgments

MSC thanks the Gates Cambridge Trust for their generous support. ACF and HRR acknowledge ERC Advanced Grant Feedback 340442. HRR also acknowledges an STFC Ernest Rutherford Fellowship. CPO and SAB acknowledge the Natural Sciences and Engineering Research Council (NSERC) for its support. BRM thanks the NSERC and the Canadian Space Agency for generous support. ACE acknowledges support from STFC grant ST/P00541/1. Support for this work was provided by the National Aeronautics and Space Administration through Chandra Award Number G05-16130X issued by the Chandra X-ray Center, which is operated by the Smithsonian Astrophysical Observatory for and on behalf of the National Aeronautics Space Administration under contract NAS8-03060.



\appendix

\section{Analytic Projection of 3D Broken Power-Law Densities}
\label{app:analytic}

To model an abrupt, spherical jump in density, one may want to implement a simple broken power law of the form:
\[ n_e(r) =
  \begin{cases}
    A_1 r^{\alpha _1}       ,& \quad  0 \leq r \leq r_c \\
    A_2 r^{\alpha _2}  ,& \quad  r_c \leq r \leq R \\
    0 \quad ,& \quad  r> R
  \end{cases},
\]
where $r$ is a three-dimensional radius, $\{ A_1,A_2 \}$ are the power law normalizations which give rise to a break strength, $\{\alpha _1, \alpha _2 \} < 0$ are the power law indices, $r_c$ is the critical radius at which the density discontinuity occurs, and $R$ is some arbitrary large ``outer'' radius of the ICM where the density goes to zero.

Then, projecting along the line of sight via the emission integral, one can find the surface brightness:
\begin{align*}
SB(x) &= 2 \int_0^{\sqrt{R^2 - x^2}} n_e n_p dl \cdot dA \\
\quad &= \left\{2 \int_x^R \frac{n_e(r) n_p(r) r dr}{\sqrt{r^2 - x^2}} \right\}\cdot dA 
\end{align*}
where $x$ is the two-dimensional radius on the plane of the sky and $dl$ is the distance along the line of sight. The integral in brackets can be done analytically, resulting in: 
\begin{equation}
\label{eq:analytic}
\begin{cases}
		A_2^2 x^{-2a_2} \left[ B\left(\frac{x^2}{r_c^2},a_2,\frac{1}{2}\right) - B\left(\frac{x^2}{R^2},a_2,\frac{1}{2}\right) \right] + A_1^2 x^{-2a_1} \left[ B\left( 1,a_1,\frac{1}{2} \right) - B\left( \frac{x^2}{r_c^2},a_1,\frac{1}{2} \right)  \right],& r<r_c\\
       A_2^2 x^{-2a_2} \left[ B\left( 1,a_2,\frac{1}{2} \right) - B\left( \frac{x^2}{R^2},a_2,\frac{1}{2} \right)  \right], & r>r_c
\end{cases}
\end{equation}
where $a_i := -(\frac{1}{2}+\alpha_i)$ 
and $B(z;a,b)$ is the incomplete Beta function, defined as:
\begin{equation*}
B(z;a,b) \equiv \int_0^z u^{a-1} (1-u)^{b-1} du
\end{equation*}
\autoref{eq:analytic} can subsequently be made into a solid of revolution multiplying by $2 \pi x$, and fit to the observed surface brightness profile to determine the location of density discontinuities (e.g. cold fronts or shock fronts) more efficiently than numerically projecting a density model onto the plane of the sky.



\bibliographystyle{apj}
\bibliography{A1664_ApJ_draft} 






\end{document}